\documentstyle {article}

\begin{document}

\title{\bf A Note on Cosmology and Loop Quantum Gravity
\footnote{The fact that the references may look old is due to the
fact that the present paper was completed in 1997}}

\author{Pablo Mora\\
Instituto de F\'{\i}sica, Facultad de Ciencias \\
Igu\'a 4225, Montevideo, Uruguay}

\date{\today}

\maketitle

\begin{abstract}
One possible description of the very early stages of the 
evolution of the universe is provided by chaotic inflationary
models for which the role of the phenomenological inflaton
field is played by quantum gravitational effects. We argue that
simple non perturbative calculation supporting and establishing
that kind of scenario may be possible
within the framework of loop quantum gravity. This approach 
could even lead
to testable predictions of quantum gravity (QG).\\
We explore this proposal in a model for QG, Rovelli's T-theory, 
without reaching definitive results. However the analysis shows that
our approach does not lead to difficulties of principle,
and could well work in a more realistic model.
\end{abstract}

It is widely accepted that quantum gravitational effects were
dominant in the very early universe. It is also assumed that
a future theory of QG should provide the framework to describe the
origin of the universe as well as its initial conditions (in terms 
of a definition of the kinematic and the dynamic that must also be
a part of the theory)\cite{zeldovich1,zeldovich2,linde1,vilenkin1}.
However it is usually believed that a subsequent inflationary period
would wash out any observable consequence of those QG effects.\\
The success of the chaotic inflation scenario \cite{linde2,linde3}
and the difficulties of the old \cite{guth1} and new \cite{linde4,albrech1}
inflationary models (see for instance \cite{linde5,mukhanov1,collins1,
vishnu1,dalia1,dalia2,kung1}) shift the problem of finding the physical
foundations of the phenomenological inflationary models from the 
Grand Unification scale ($\approx 10^{-5} M_{Pl}$) to the Planck
scale. And that is just the scale in which QG effects are expected 
to be important.\\
It is well known that the chaotic inflationary scenario can be realized in
a wealth of physical model, including models in which the role of the 
inflaton field is played by higher curvature terms in the Einstein-Hilbert 
action coming from quantum corrections. To this class belongs the Starobinsky
model \cite{starobinsky1}, often considered the first inflationary model,
in which inflation is driven by second order curvature terms coming
from vacuum polarization of matter fields. Starobinsky inflationary 
solution has been
shown by Simon \cite{simon} not to be valid within the 
limits of the semiclassical gravity 
perturbative approach used to obtain the higher curvature corrections
in the action of the model. However the fact remains that higher curvature 
theories are known to often have de Sitter-like solutions\cite{barrow1}.
It is known furthermore that
quantum corrections to pure gravity (vaccum polarization
of gravitons) lead to higher order curvature terms \cite{hooft1}.\\
A recent work on the above mentioned approach to inflation is 
Reference\cite{brandenberger}.\\
This kind of picture seems reasonable from general physical considerations.
Out of the different kinds of physical effects able in principle of
driving an inflationary period, the first one in taking over would
dominate, because the energy densities of the other fields will be
rapidly diluted (scalar fields energy densities would decrease to a
constant, much smaller anyway than the "inflaton" energy density), while the
contribution of the "inflaton" part would remain esentially constant.
Therefore we can argue that if QG effects can support an inflationary 
expansion, then they would be the ones to do it, because they dominate
at an earlier time (the Planck era).\\
The problem with the standard perturbative approach is that close
to the initial state the curvature and all the higher curvature
counterterms are divergent. Even if we keep the lower orders
with a suitable redefinition of the constants such an expansion around a 
smooth geometry is unlikely to be a good approximation for the possibly 
very non classical singular (zero volume) initial state.\\
An alternative approach could be provided by the loop quantum gravity 
program \cite{rovelli1}. In that framework the eigenstates and eigenvalues
of the volume operator are known \cite{rovelli2,thiemann1}, including a 
sector of states with zero volume (described in a completely non singular 
way). The space of states is actually simpler for zero or small volumes.
The time evolution of the physical states starting in some zero volume
initial state\footnote{our opinions about how this initial state could be determined
and about how a time variable ought be chosen in quantum gravity are stated
elsewhere \cite{doldan1}. See also references therein for other approaches
to the problem of the initial conditions and to the "issue of time"}
can in principle be adressed in a non perturbative 
way (see \cite{rovelli3} and references therein) without any infinities.
From there one could compute the evolution of the expectation value of the 
volume operator $\langle V\rangle (t)$ (and $\Delta V(t)$)
\footnote{We could have otherwise a set of quasi classical 
decohering histories with $V_1(t)$, $V_2(t)$,....The role of
the "enviroment" could be played by the degrees of freedom of loop
QG that are
non zero for zero volume states, which should be traced out}. 
If $\langle V\rangle (t)$ were an exponential (or some other rapidly increasing
function) and 
$\frac{\Delta V(t)}{\langle V\rangle (t)}$  were small we could argue 
that we had a quasi classical evolution and that the irregularities
would be smoothed away by the fast expansion, giving a 
Friedmann-Robertson-Walker (FRW) metric with scale factor 
$a(t)$ proportional to $V^{1/3}(t)$. The knowledge of $a(t)$ is
all what is required to be able
to make concrete predictions about the spectrum of primordial
density fluctuations, or the spectrum of fossil gravitational waves,
using standard techniques (see for instance \cite{linde3}).\\
A recent loop quantum gravity model that allows concrete
computations of the evolution of physical states, expected values
of operators and transition amplitudes is provided by Rovelli's T-theory
\cite{rovelli3}.
This model consists of of standard General Relativity (GR) with a negative
cosmological constant and a massless scalar field which is to play
the role of a "clock field" after some gauge fixing.The action of the 
model is

\begin{eqnarray}
S[g_{\mu\nu},\phi]=-\int d^4x\sqrt{\mid g\mid}[\frac{R-\lambda}
{16\pi G}+\frac{1}{2}g^{\mu\nu}\partial _{\mu}\phi\partial _{\nu}\phi]
\end{eqnarray}

giving the classical equations of motion

\begin{eqnarray}
R_{\mu\nu}-\frac{1}{2}g_{\mu\nu}R+\frac{1}{2}g_{\mu\nu}\lambda=
-8\pi G[\partial _{\mu}\phi\partial _{\nu}\phi -\frac{1}{2}g_{\mu\nu}
\partial _{\mu}\phi\partial ^{\mu}\phi]\\
D^{\mu}D_{\mu}\phi=0
\end{eqnarray}

In order to do the canonical analysis of the model, the gauge freedom
of the model (general covariance) is fixed by the gauge (coordinate)
conditions

\begin{eqnarray}
\phi ({\bf x},t)=\mu T\\
N_i=g_{i0}=0
\end{eqnarray}

Notice that choice of the time variable $T$ is only possible
if the hypersurfaces $\phi ({\bf x},t)=constant$ are spacelike,
what is not true in general (for instance 
for a stationary solution they would
be timelike). We will restrict ourselves to the situations in which
this gauge fixing is possible. This clearly might be a problem at 
the quantum 
level for wich all configurations should in principle be included.\\
The canonical analysis gives then the true hamiltonian

\begin{eqnarray}
H=\sqrt{2}\mu\int d^3 x \sqrt{q}\sqrt{\rho _{\lambda}-{\cal C}_{ADM}}
\end{eqnarray}

where $q$ is the determinant of the 3-metric and 
${\cal C}_{ADM}$ is the usual ADM hamiltonian constraint. For $H$
to be a well defined real hamiltonian it must be 
$\rho _{\lambda}-{\cal C}_{ADM}>0$.\\
The following step is to chose the initial condition in such a way that
the negative energy density of the vacuum due to the cosmological
constant term cancels the positive density of the "clock field". This
is called the "well balanced clock regime" (WBCR) and it can hold
approximately only for a while. We have in the WBCR

\begin{eqnarray}
{\cal C}_{ADM}=\rho _{TOT}=\rho _{\phi}-\rho _{\lambda}=0
\end{eqnarray}

We ask furthermore the lapse function to be equal to one 
initially $N(0)=1$, what imply that, for small $T$, $T$ is
approximately the proper time. But

\begin{eqnarray}
\rho _{\phi}=\frac{1}{2N}\frac{\partial\phi}{\partial T}
\frac{\partial\phi}{\partial T}=\frac{1}{2N}\mu ^2
\end{eqnarray}

Then from the WBCR and $N(0)=1$ it follows

\begin{eqnarray}
\frac{\lambda}{16\pi G}=\rho _{\lambda}=\frac{\mu ^2}{2}
\end{eqnarray}
which forces a choice of $\mu$ given $\rho _{\lambda}$.
The hamiltonian becomes

\begin{eqnarray}
H=2\rho _{\lambda}\int d^3 x \sqrt{q}
\sqrt{1-\frac{{\cal C}_{ADM}}{\rho _{\lambda}}}
\end{eqnarray}

Now we will look for FRW solutions for the model to see how
does it work at the level of classical cosmology. The Friedmann
equations for the model are

\begin{eqnarray}
\ddot{\phi}+3\frac{\dot{a}}{a}\dot{\phi}=0\\
(\frac{\dot{a}}{a})^2+\frac{k}{a^2}=\frac{4\pi G}{3}
[\frac{1}{2}\dot{\phi}^2-\rho _{\lambda}]\\
\ddot{a}=-\frac{8\pi G}{3}[\dot{\phi}^2+\rho _{\lambda}]
\end{eqnarray}

where the dots represent derivatives with respect to the proper
time $t$.
Notice that the energy density $\rho$ and the pressure $p$
are given by

\begin{eqnarray}
\rho =\frac{1}{2}\dot{\phi}^2-\rho _{\lambda}\\
p=\frac{1}{2}\dot{\phi}^2+\rho _{\lambda}
\end{eqnarray}

from where it follows that even in the WBCR the pressure will
be non zero.\\
We have to first order in t\\

\begin{tabular}{|c|c|c|}
\hline
{$k$}&{$a(t)$}&{$\phi (t)$}\\\hline
{+1}&{$a_0+it$ (unphysical)}&\\\hline
{0}&{$a_0+{\cal O}(t^2)$}&{$\phi _0+\dot{\phi _0}t$}\\\hline
{-1}&{$a_0\pm t$}&{$\phi _0+\dot{\phi _0}t$}\\\hline
\end{tabular}
\\
For $k=0$ we have the exact solutions

\begin{eqnarray}
\phi -\phi _0=\pm\frac{1}{\sqrt{6\pi G}}ln\mid 
sec[\sqrt{12\pi G\rho_{\lambda}}(t-t_0)+C_{\pm}]+
tan[\sqrt{12\pi G\rho_{\lambda}}(t-t_0)+C_{\pm}]\mid\\
a=a_0\{ cos[\sqrt{12\pi G\rho_{\lambda}}(t-t_0)+C_{\pm}]\} ^{\frac{1}{3}}
\end{eqnarray}

where $C_{\pm}=sec^{-1}(\pm\frac{\dot{\phi}_0}{\sqrt{2\rho _{\lambda}}})$. In the
WBCR and for $\dot{\phi}>0$ in a neighborhood of $t=0$ we get

\begin{eqnarray}
\phi -\phi _0=\frac{1}{\sqrt{6\pi G}}ln\mid 
sec[\sqrt{12\pi G\rho _{\lambda}}(t-t_0)]+
tan[\sqrt{12\pi G\rho_{\lambda}}(t-t_0)]\mid\\
a=a_0\{ cos[\sqrt{12\pi G\rho_{\lambda}}(t-t_0)]\} ^{\frac{1}{3}}
\end{eqnarray}

or, for $\phi _0=0$ and $\phi =\mu T$

\begin{eqnarray}
a/a_0=\frac{4}{[cosh(3\mu \sqrt{6 \pi G}T)+3cosh(\mu\sqrt{6\pi G} T)]}
\end{eqnarray}

Notice that $a$ collapses to 0 while $\phi$ diverges as functions of $t$ for
$t-t_0=\frac{1}{4}\sqrt{\frac{\pi}{3}}\frac{1}{\sqrt{G \rho_{\lambda}}}$,
but $a(T)$ is well behaved for $T>0$. For large $T$ the radius $a$
is proportional to $e^{-3\mu \sqrt{6 \pi G}T}$

At the quantum level we will consider the case in which the matrix
elements of $\frac{{\cal C}_{ADM}}{\rho _{\lambda}}$ in some basis
$\mid s_a\rangle$ to be specified below satisfy

\begin{eqnarray}
\langle s_a\mid\frac{{\cal C}_{ADM}}{\rho _{\lambda}}\mid s_b\rangle \ll 1
\end{eqnarray}

what is just the quantum version of the WBCR. 
In that case the hamiltonian operator can expanded\footnote{Note that
is not immediate that the hamiltonian is well defined and hermitian, what
would require the operator $1-{\cal C_{ADM}}/\rho _{\lambda}$ to be 
positive definite and hermitian. See \cite{rovelli3} about this point} 
as

\begin{eqnarray}
H=2\rho _{\lambda}\int d^3 x \sqrt{q}
\sqrt{1-\frac{{\cal C}_{ADM}}{\rho _{\lambda}}}
\approx \\\nonumber
2\rho _{\lambda} V+\int d^3 x \sqrt{q}
\sqrt{{\cal C}_{ADM}}=H_0+W
\end{eqnarray}

where $V$ is the volume operator and

\begin{eqnarray}
H_0=2\rho _{\lambda} V\\
W=\int d^3 x \sqrt{q}\sqrt{{\cal C}_{ADM}}
\end{eqnarray}

It is plausible that the WBCR and the parameter choice
$\rho _{\lambda}=\frac{\mu ^2}{2}$ in the quantum case, plus
the correspondence principle would yield that initially $T$ is
approximately the proper time.

The base of eigenstates of the unperturbed hamiltonian is 
the base of eigenstates of the volume operator. That is the
so called s-knot states basis \cite{rovelli1,rovelli2,thiemann1,rovellivol},
satisfying

\begin{eqnarray}
\langle s\mid s'\rangle =\delta _{ss'}\\
V\mid s\rangle =V(s)\mid s\rangle
\end{eqnarray}

where $V(s)$ is the eigenvalue of the state. The spectrum of the volume
operator is believed by the experts to be discrete, though
a mathematical proof of the absence of accumulation points in it is
still missing \cite{thiemann1}.
The evolution of the s-knot states in the unperturbed theory is

\begin{eqnarray}
\mid s,t\rangle =e^{-iE_sT/\hbar}\mid s\rangle
\end{eqnarray}

with

\begin{eqnarray}
E_s=2\rho _{\lambda}V(s)
\end{eqnarray}

A perturbative approach is valid for 
$M_{Pl}\ll\rho _{\lambda}V(s)$ because

\begin{eqnarray}
W_{ab}=\langle s_a\mid W\mid s_b\rangle=M_{Pl}{\cal M}_{ab}
\end{eqnarray}

with ${\cal M}_{ab}\approx 1$ for $a\ne b$ and for the lower volume 
states \footnote{This follows
from the action of ${\cal C}_{ADM}$ on spin network states as given
by Borissov \cite{borissov1}. However the action of the hamiltonian
constraint on spin network states is still matter of debate
\cite{thiemann2,smolinh}}.
Standard perturbation theory gives the probability amplitudes, 
given an initial state $\mid s_i\rangle$, as

\begin{eqnarray}
\langle s_f,T\mid s_i,0\rangle =W_{fi}
\frac{1-e^{i\frac{(E_f-E_i)}{\hbar}T}}{E_f-E_i}+.....
\end{eqnarray}

and transition probabilities

\begin{eqnarray}
{\cal P}_{fi}(T) =\mid W_{fi}\mid ^2
\frac{1-cos[\frac{(E_f-E_i)}{\hbar}T]}{(E_f-E_i)^2}+.....
\end{eqnarray}

These expressions are good approximations for ${\cal P}_{fi}(T)\ll 1$.
The transition probabilities are strongly peaked around 
$E_f=E_i$. That means that transitions between states with
different volume are strongly supressed. For instance for
$\rho _{\lambda}=100$ and  $\Delta V=1$ we get 
${\cal P}_{fi}(T)<\frac{1}{5000}$. This could be interpreted as
a prediction of freezing on the evolution of the volume of the 
universe, however the situation is not so simple. The reason
is that the spacing between states with different values of the
volume decreases with the volume (or equivalently the density of
states increases) faster than $1/V$
\footnote{To see this is enough to notice that for a class of s-knots 
with one four valent vertex the eigenvalue of the volume operator
is $2 L_{Pl}^3\sqrt{\mid \sigma\mid}
[j_1j_2j_3(j_1+j_2+j_3)]^{\frac{1}{4}}$ where $\sqrt{\mid \sigma\mid}$
is a number of order one and the $j$'s are arbitrary positive 
integers or positive half integers and $L_{Pl}$ is the Plank
length \cite{thiemann1}},
then though for 
the lower volume
states (say smaller than 10 in Planck units) which have volume differences
of order one we do have a frozen volume situation, after enough time
the system could reach a higher volume state for which there are many 
states with very close values of the volume. To know what would happen
in that case would require higher orders in perturbation theory as well
as a detailed knowledge of the $W_{ab}$'s\footnote{
An important point is that because for vertices of valence four or
higher there is a continuous of s-knot states \cite{rovelliknot}, the
operator $W$ could induce transitions between subspaces corresponding
to different volumes with very different dimensions, what could
make transitions almost certain. However if we accept Borissov action
for the hamiltonian constraint \cite{borissov1} we can rule out that
possibility, because it only adds trivalent vertices.}
. If the volume is actually increasing
that would contrast with the classical behaviour for a FRW metric with
$k=0$.\\
We have not been able to actually compute the evolution of the volume.
Despite of that the study of the model shows that the kind of calculations
suggested,
giving a quantum gravitational prediction for $V(t)$ from 
some zero volume initial state
(which might be uniquely determined from other considerations) are
already possible. And, as we pointed out before, that calculation may
lead to experimentally testable predictions of quantum gravity.\\

I would like to thank T. Jacobson for disscussions on the topics of this
paper, even though he would probably would not agree with the approach 
followed here. As I said before this paper was completed in 1997, and
at this point I am myself have became quite doubtful about 
several points of such approach.
I am publishing it here anyway because I believe some people may still find some 
points of interest in it.

\end{document}